\documentclass[pra,aps,twocolumn,groupaddresses,superscriptaddress]{revtex4-1}
\usepackage[colorlinks=true, allcolors=blue]{hyperref}
\usepackage{graphicx}
\usepackage{amsmath,amssymb,bm}
\usepackage{mathtools}
\usepackage{latexsym}
\usepackage{verbatim}
\usepackage{times}
\usepackage{etoolbox}

\newcommand{\abs}[1]{\left\vert#1\right\vert}

\newcommand{\ket}[1]{\left\vert#1\right\rangle}
\newcommand{\bra}[1]{\left\langle#1\right\vert}

\newcommand{\mb}[1]{{\bm#1}}
\newcommand{\tr}{\operatorname{Tr}}

\newcommand{\beq}{\begin{equation}}
\newcommand{\eeq}{\end{equation}}
\newcommand{\baq}{\begin{eqnarray}}
\newcommand{\eaq}{\end{eqnarray}}

\newcommand{\brac}[1]{\left\lbrace #1
	 \right\rbrace}
\newcommand{\intinf}{\int_{-\infty}^\infty dx}

\def\ket#1{| #1 \rangle}
\def\bra#1{\langle #1 |}

\def\ave#1{\langle #1 \rangle}

\def\Tr{\operatorname{Tr}}

\newcommand{\sbrac}[1]{\left[#1\right]}
\newcommand{\pbrac}[1]{\left(#1\right)}

\def\F{\mathcal{F}}

\begin{document}
	
	\title{Quantum precision of beam pointing}
	\author{Haoyu Qi}
	\email{haoyu@xanadu.ai}
	\author{Kamil Br{\'a}dler}
	\author{Christian Weedbrook}
	\affiliation{Xanadu, Toronto, ON, M5G 2C8, Canada}
	\author{Saikat Guha}
	\affiliation{College of Optical Sciences, University of Arizona,
		1630 E. University Blvd., Tucson, Arizona 85719, USA}

\begin{abstract}
We consider estimating a small transverse displacement of an optical beam over a line-of-sight propagation path: a problem that has numerous important applications ranging from establishing a lasercom link, single-molecule tracking, guided munition, to atomic force microscopy. We establish the ultimate quantum limit of the accuracy of sensing a beam displacement, and quantify the classical-quantum gap. Further, using normal-mode decomposition of the Fresnel propagation kernel, and insights from recent work on entanglement-assisted sensing, we find a near-term realizable multi-spatio-temporal-mode continuous-variable entangled-state probe and a receiver design, which attains the quantum precision limit. We find a Heisenberg-limited sensitivity enhancement in terms of the number of entangled temporal modes, and a curious super-Heisenberg quantum enhanced scaling in terms of the number of entangled spatial modes  permitted by the diffraction-limited beam propagation geometry.
\end{abstract}

\maketitle

\section{Introduction}

The precision of optical sensors of both {\em active} (e.g., laser gyroscopes \cite{lee2003review}, LIDARs \cite{amann2001laser}, atomic-force microscopes \cite{eaton2010atomic}, and laser vibrometers \cite{castellini2006laser}), and {\em passive} (e.g., fluorescence microscopy \cite{lichtman2005fluorescence}, astronomical imaging \cite{howell2006handbook}, and satellite based remote sensing \cite{degnan1993millimeter}) kinds is often quantified as the standard deviation $\delta \theta$ of the estimate of desired scene parameter(s) $\theta$ versus the total mean photon number (a.k.a. power) $N$ collected over the receiver's integration time. The fundamental precision limit, i.e., the best scaling of $\delta \theta$ versus $N$ achievable by using the optimal probe light and the receiver, given the physical constraints of the application scenario, is ultimately governed by quantum mechanics.



When multiple sensors have different {\em views} of the same scene, pre-shared entanglement across those sensors can improve the attainable precision. This is true both for passive~\cite{gottesman2012longer} and active \cite{giovannetti2006quantum} sensors. In recent years, several theoretical calculations \cite{humphreys2013quantum,proctor2018multiparameter,zhuang2018distributed} (for active sensing) have indicated that if a set of $M$ distributed sensors are sensing one global parameter of the scene, then pre-shared entanglement among the sensors can help improve the sensing precision. As an example, for sensing an average phase modulation across $M$ sensors, for a total of $N$ probe photons expended across $M$ distributed sensors, individual (non-entangled) quantum sensors obtain a standard deviation $\delta \theta \sim M^{3/2}/N$. Whereas, a probe entangled across those $M$ sensors yields $\delta \theta \sim M/N$~\cite{humphreys2013quantum}. So, for this problem, shared entanglement improves the sensing precision by a factor of $\sqrt{M}$. Another example is when supplying entanglement to a large sensor array of a radio-frequency (RF) photonic receiver can get a tremendous boost in sensing a single parameter, such as the angle of incidence of the received RF field~\cite{Xia2019-fp}.

\begin{figure}[!htbp]
	\centering
	\includegraphics[width=0.8\columnwidth]{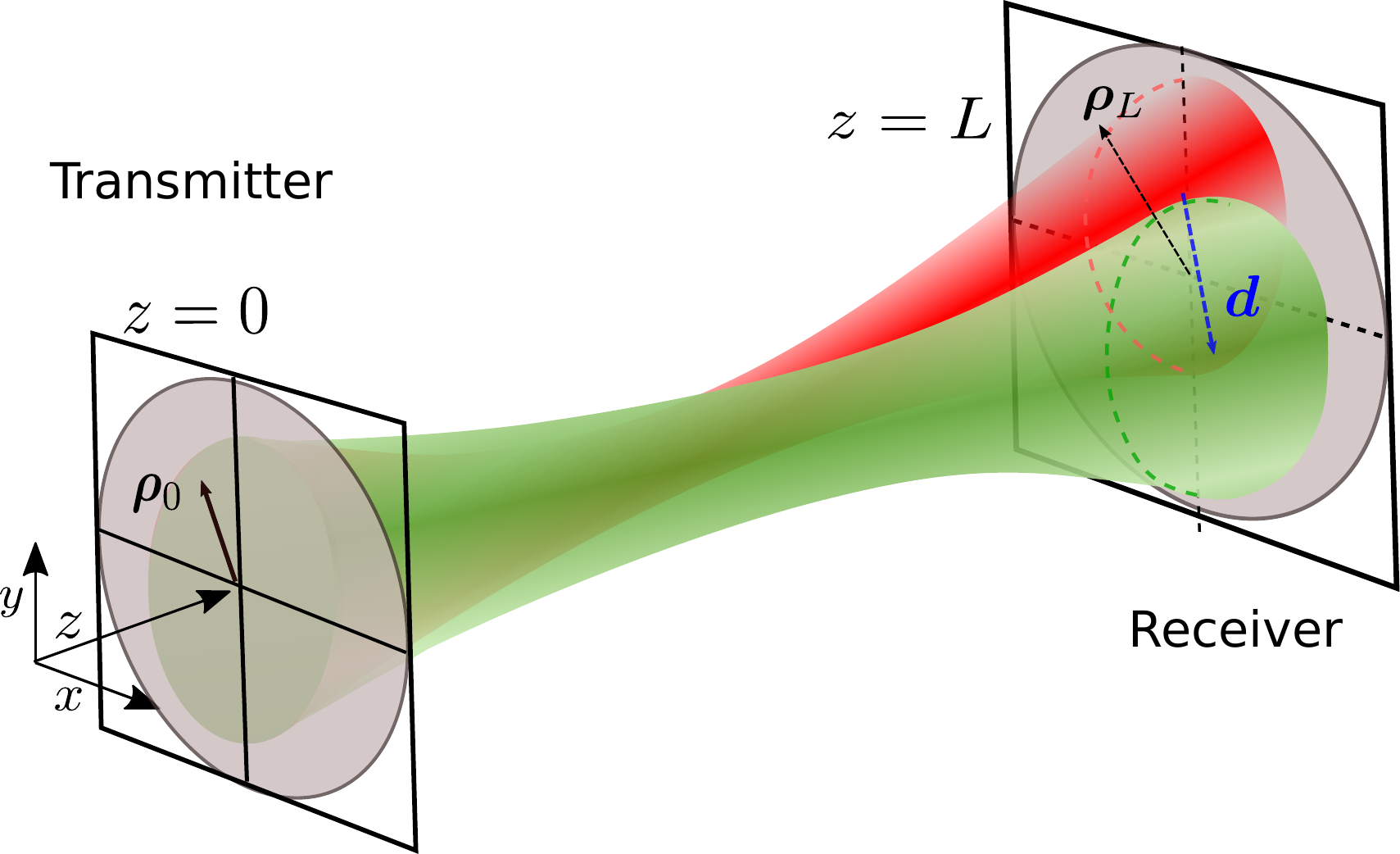}
	\caption{An $L$-meter line-of-sight optical propagation path with circular transmitter and receiver pupils of areas $A_t$ and $A_r$, respectively. We consider the near-field propagation regime, i.e., $D \equiv A_tA_r/(\lambda L)^2 \gg 1$, where $\lambda$ is the center wavelength of the transmitted light. A transmitter of $W$ [Hz] optical bandwidth modulates $M_S \approx D$ near-unity-transmissivity orthogonal spatial modes and $M_T \approx WT$ orthogonal temporal modes over probing duration $T$ seconds, with a total of $N$ mean photon number distributed over $M = M_SM_T$ modes. The transmitter points the beam towards the center of the receiver aperture. However, there is an unknown transverse (vector) displacement of the beam, $\mb{d}$, measured in the receiver-aperture plane, whose origin is dependent on the application. The goal of the receiver---via optimal detection of the collected light---is to estimate $\mb{d}$ precisely. In this paper, we will restrict our attention to a scalar beam displacement $d$ along the $x$ axis.}
	\label{fig:setup}
\end{figure}

One does not need $M$ physical sensors to see the aforesaid entanglement-assisted performance improvement. As we show in this paper, entanglement across multiple orthogonal spatio-temporal modes of a probe field---each of which are non-trivially, independently, modulated by the target parameter of interest (and hence these modes can be thought of as ``multiple sensors")---can improve the performance of a standalone active sensor. We will show such performance improvement in the accuracy of detecting a small transverse displacement of an optical beam over a near-field free-space propagation path (see Fig.~\ref{fig:setup}).

The paper is organized as follows. In Section~\ref{sec:statement}, we describe the problem of beam displacement sensing, and contextualize our contributions with prior work in this topic. In Section~{sec:synopsis}, we provide a brief synopsis of known categories of quantum enhancements in optical sensing problems. Section~\ref{sec:results} defines all the important parameters describing the problem setup, and states the main results of the paper, without proof.
In Section~{sec:modeling}, we develop the underlying free-space propagation model, and show how a small transverse beam displacement manifests itself as an array of phase modulations on a set of orthogonal modes, interspersed with linear mode transformations. In Section~\ref{sec:classical}, we derive the performance of the optimal classical probe, i.e., when all the spatial and temporal modes are excited in a coherent state. In Section~\ref{sec:multimode}, we derive the performance of the optimal quantum probe, and find a multimode squeezed-light probe that attains it. Section~{sec:structured} describes a structured transmitter-receiver design to implement the aforesaid squeezed-light probe and multimode homodyne detection, which would attain the quantum-limited performance. Section~\ref{sec:conclusion} concludes the paper, with a discussion of fruitful future extensions of this work.

\section{Problem statement and background}\label{sec:statement}

\noindent Accurate estimation of a small transverse displacement of an optical beam is important in many practical applications. Examples include ultra-stable beam pointing, acquisition and tracking for space-based laser communications, in LIDAR receivers, for precision guided munition, ultra-weak absorption measurements, single-molecule tracking \cite{taylor2013biological,taylor2014subdiffraction} in biological imaging and atomic force microscopy, just to name a few. Our goal in this paper is to study the fundamental precision limit of estimating a small transverse displacement $d$ of an optical beam over a near-field free-space propagation path shown in Fig.~\ref{fig:setup}, and quantify the performance gap between the best classical optical probe and the optimal quantum entangled probe.

The first paper that tackled this problem~\cite{fabre_quantum_2000} considered a {\em split detector}---one which has two pixels separated by an edge---and claimed that no single-mode probe state can surpass the standard quantum limit (SQL) or the so-called {\em shot noise scaling} of measurement sensitivity. This led the authors to consider multi-spatial-mode probe states, and in particular a two-mode entangled state prepared by interferometrically mixing ideal laser light (coherent state) and squeezed light. They showed that the error of estimating the transverse beam displacement was $\sim 1/(\sqrt{N} e^r)$, where $r$ quantifies the amount of squeezing and $N$ is the total mean photon number expended during the probing interval. This idea was implemented in a proof-of-concept experiment in \cite{treps2002surpassing}. In Ref.~\cite{treps2003quantum}, this idea, dubbed the ``quantum laser pointer", was generalized to two dimensions and an experiment was carried out. 

There were several follow up papers in terms of a more theoretical analysis for this problem. For example, Ref.~\cite{hsu_optimal_2004} considered a split detector, used a squeezed light probe, and compared the performance with a photon-number-resolving (PNR) detector array as a baseline. In Ref.~\cite{delaubert2008quantum}, the authors calculated the classical Fisher information (CFI)~\cite{VanTrees} of the PNR array and that of a split detector, but from an imaging point of view. The quantum Fisher information (QFI)~\cite{Helstrom1969-lt,braunstein1994statistical,braunstein1996generalized} for this problem---a measure of the optimal sensitivity in detecting a parameter embedded in a quantum state with no restrictions imposed on how the quantum state is detected---was considered in Ref.~\cite{pinel2012ultimate}. The problem they considered, that of estimating an arbitrary parameter encoded in a multi-mode Gaussian quantum state, is slightly more general than the problem we consider here, where that parameter is a transverse displacement of a beam. Their conclusion was that the optimal Gaussian state is a coherent state combined with a single-mode squeezed vacuum. However, no statement on Heisenberg limited sensitivity was made, non-Gaussian state probes were not considered, and most importantly, no structured receiver design was proposed or analyzed. 

In all of the papers discussed above, the authors assumed a strong coherent state probe, modulated with quadrature-squeezed light. Finding the quantum optimal (potentially spatio-temporal-entangled) probe, its performance, the role of entanglement in space versus time, and structured transmitter-receiver designs to harness this ultimate quantum enhancement in the precision of estimating beam displacement were left open. This paper addresses all of the above.


\section{Brief synopsis of optical quantum sensing}\label{sec:synopsis}
All known quantum enhancements in optical sensing applications can be broadly classified as follows:
\begin{enumerate}
\item {\em Classical probe with a quantum-enhanced receiver}---Example applications of this lie both in: {\em (a) standoff sensing}: a lidar with an inefficient heterodyne receiver can get a performance boost by preceding the receiver with a quantum-noise-limited phase-sensitive amplifier (squeezer)~\cite{Dutton2010-fj}; and {\em (b) near-field sensing}: optical reading of a passively-encoded memory, such as a DVD, using a laser probe can benefit from a quantum-optical joint-detection receiver, especially in the low-photon-flux operation regime~\cite{Guha2013-ap}.

\item {\em Quantum probe with a quantum receiver}---This paper's application, and the majority of papers in quantum metrology, fall in this category~\cite{Giovannetti2011-cw}. Non-classical light is sent, in its entirety, to a target, and upon modulation of the light by the target, a quantum receiver is used to extract the target parameter of interest. This kind of sensor shows the so-called Heisenberg limited sensitivity: the standard deviation of the parameter estimate scales as the inverse, as opposed to the classical scaling of inverse of the square root, of the total expended energy. The Heisenberg scaling disappears even with the slightest amount of loss or noise, but there can be a large constant-factor enhancement in the inverse-square-root scaling of the estimate's standard deviation, depending upon how much loss there is in the sensing channel. When the target-modulated light accrues high losses, this class of sensing does not afford much quantum advantage. This class of sensing further benefits from entangling multiple sensors when sensing a global parameter of interest~\cite{humphreys2013quantum,proctor2018multiparameter,zhuang2018distributed}. Quite well-known in this sensor category are N00N state~\cite{Dowling2008-wb}, constant-photon-number~\cite{Xiang2013-ll} and squeezed-light probes~\cite{Lawrie2019-fo}.

\item {\em Quantum illumination with an entangled reference beam}---The aforesaid no-go result for quantum-enhanced sensing in high losses does not hold when there is a high (thermal) noise in the channel, relevant, e.g., in the microwave regime. In this regime, a quantum probe where the transmitter retains a reference beam entangled with the transmitted signal~\cite{Tan2008-pa}, and uses a quantum receiver~\cite{Guha2009-rw} to jointly detect the target-return light with the retained reference, can yield improved performance over any sensor that employs a classical-light probe. Quantum illumination is a peculiar quantum sensing paradigm where the original entanglement gets broken (by the lossy-noisy channel) in the act of giving the sensor its quantum enhancement. 

\item {\em Quantum limited classical passive sensing}---In recent years, quantum estimation tools have been employed to evaluate fundamental limits of passive imaging, with applications for example to astronomy and fluorescence microscopy~\cite{Tsang2016-lq,Dutton2019-ig,Grace2019-ei}. The receivers that attain quantum limited performance involve pre-detection linear mode transformations and photon detection, and hence completely describable by the semi-classical (shot-noise) theory of photodetection. So, these sensors are completely classical. But, quantum treatment of the scene-irradiated light lets us use powerful quantum-estimation tools without which finding the optimal receiver designs would be very hard, if not impossible.

\end{enumerate}

\section{Main Results}\label{sec:results}

Let us consider a quasi-monochromatic optical probe of linewidth $W$ Hz around center-wavelength $\lambda$, and a near-field $L$-meter-range line-of-sight propagation geometry shown in Fig.~\ref{fig:setup}. The Fresnel number product $D \equiv A_tA_r/(\lambda L)^2 \gg 1$, where $A_t$ and $A_r$ are areas of the transmitter and receiver apertures. There are roughly $M_S \approx D$ near-unity-transmissivity orthogonal spatial modes and $M_T \approx WT$ orthogonal temporal modes over a probing duration of $T$ seconds. The source points the beam towards the center of the receiver aperture. However, there is an unknown transverse displacement of the beam, $d$, measured in the plane of the receiver aperture's entrance pupil, whose origin is dependent on the application. The goal of the receiver---via optimal detection of the collected light---is to estimate $d$ precisely.

Let us impose a transmit power constraint of ${\bar n}$ mean photon number per mode, distributed over the $M = M_SM_T$ spatio-temporal modes. This implies a total of $N = {\bar n}M_SM_T$ mean photon number over the probing duration, and equivalently a transmit power constraint of $P = {\bar n}M_S W$ photons per second. Note that power in Watts would be $P (hc/\lambda)$, where $hc/\lambda$ is the photon energy at wavelength $\lambda$.

We find the following main results for scaling (constants omitted) of the standard deviation $\delta d$ of the beam displacement estimate.

\begin{enumerate}
	
	\item {\textbf{Optimal classical probe.}} If the transmitted light is constrained to be {\em classical}, in other words expressible as a statistical mixture of coherent states of the $M$ spatio-temporal modes (i.e., have a proper $P$-function representation),
	
	\begin{eqnarray}
	\delta d \sim \frac{1}{\left(\sqrt{M_T}M_S\sqrt{\bar n}\right)} = {\frac{1}{\sqrt{M_S}}} \times \frac{1}{\sqrt{PT}};
	\label{eq:scaling1}
	\end{eqnarray} 
	
	\item {\textbf{Optimal spatially-entangled probe.}} If we allow the probe to be entangled over all $M_S$ spatial modes, but there is no entanglement (i.e., product state) across the $M_T$ temporal modes, then we have, with the optimal quantum probe:
	
	\begin{eqnarray}
	\delta d \sim \frac{1}{(\sqrt{M_T}M_S^{3/2}{\bar n})} = {\frac{W}{\sqrt{M_S}}} \times \frac{1}{P\sqrt{T}};
	\label{eq:scaling2}
	\end{eqnarray}
	
	\item {\textbf{Optimal spatio-temporally entangled probe.}} If the optical probe is allowed to be entangled across all $M_S$ spatial modes and $M_T$ temporal modes, we have:
	
	\begin{eqnarray}
	\delta d \sim \frac{1}{({M_T}M_S^{3/2}{\bar n})} = {\frac{1}{\sqrt{M_S}}} \times \frac{1}{PT}.
	\label{eq:scaling3}
	\end{eqnarray}
	
\end{enumerate}

We expressed $\delta d$ above in two equivalent forms. The first form shows how $\delta d$ scales differently with an increasing number of spatial ($M_S$) and temporal ($M_T$) modes, or degrees of freedoms, respectively; as well as with respect to the mean photon number per mode, $\bar n$. This mathematical form of scaling is more readily relatable to the existing literature on quantum metrology. One sees that even with a probe entangled over multiple spatial modes (but not across temporal modes), one gets the $\delta d \sim 1/{\bar n}$ scaling, commonly known as Heisenberg limited (HL) sensitivity, as opposed to $\delta d \sim 1/\sqrt{\bar n}$, commonly known as the standard quantum limited (SQL) sensitivity of a classical sensor. However, in addition to this Heisenberg limited sensitivity in $\bar n$, we see how the $\delta d$ scales in the number of entangled spatial modes ($1/M_S \to 1/M_S^{3/2}$) and the number of entangled temporal modes ($1/M_T^{1/2} \to 1/M_T$). In this problem, we see an unconventional quantum improvement in estimation precision with respect to the number of spatial modes. This has to do with a subtlety with regards to how the beam displacement appears as a progressively higher phase modulation in an effective Mach-Zehnder array representation of the modal modulation caused by beam displacement, as the entanglement shifts to higher-order spatial modes (see Fig.~\ref{fig:unitary}). 

The second form in which we show the scaling of $\delta d$ for the three cases above is more operational. The number of near-unity-transmissivity spatial modes $M_S$ is a fixed parameter determined by the channel geometry, so we treat it as a constant. Similarly, the center wavelength $\lambda$ and the total optical bandwidth around it $W$ are treated as given. The user controlled parameters are the transmit power $P$ and the interrogation time $T$, where $PT$ is the total energy. For a classical sensor, $\delta d \sim 1/\sqrt{PT}$ (SQL), whereas for the optimal spatio-temporally entangled sensor, $\delta d \sim 1/{PT}$ (HL). A probe that is only entangled in spatial modes but not in temporal modes achieves an intermediate precision, $\delta d \sim 1/P\sqrt{T}$.

In addition to finding the performance of optimal classical and quantum sources, we propose an explicit transceiver design that achieves the optimal quantum scaling of $\delta d$ using a multi-mode-entangled squeezed-light probe and a multi-mode coherent-detection optical receiver.

Note that all the results stated above assume no loss over the propagation path. This is justified since the $M_S \approx D$ orthogonal modes have near-unity power transmissivity over a near-field propagation path ($D \gg 1$). For example, with $\lambda = 1550$nm, $10$cm radii apertures, and $L=1$km, we get $D \approx 410$. However, despite no loss due to diffraction, in any realistic application, there will be losses due to multiple factors depending upon the scenario, including: atmospheric scattering and turbulence, scattering from a quasi-lambertian reflection from a rough target surface, coupling inefficiency of the received light into the receiver, and the detector's intrinsic inefficiencies. When optical loss is included in the analysis, the Heisenberg and super-Heisenberg scalings (with respect to $M_S$ and $M_T$) will go away, and $\delta d$ will scale just like that of the optimal classical probe. However, there will be a constant factor improvement in $\delta d$ in the long integration time limit, which can be significant (e.g., an order of magnitude or more) if the losses are moderate. We leave a detailed analysis of this problem, with losses included, for future work.

\section{Quantum modeling of the problem}\label{sec:modeling}
Consider a line-of-sight free-space diffraction-limited optical transmission setup between two circular-shaped transmitter and receiver apertures with radii $r_T$ and $r_R$ respectively, as shown in Fig.~\ref{fig:setup}. An optical source at the transmitter produces a quasi-monochromatic quantum field $\hat{E}(\mb{r},t)$ of center wavelength $\lambda$ and optical bandwidth $W$, spatially limited to the exit aperture of the transmitter pupil, $\brac{\mb{\rho_0}: \abs{\mb{\rho_0}}\leq r_T}$, and temporally limited to the interval $\brac{t:t_0-T\leq t \leq t_0}$. We use $\mb{r} = (x,y,z)$ for 3D spatial coordinates, and $\mb{\rho_u} = (x,y)$ for the transverse spatial coordinates at $z = u$. After propagating through $L$ meters along the $z$ direction, the field is collected by the entrance pupil of the receiver aperture, $\brac{\mb{\rho}_L: \abs{\mb{\rho}_L}\leq r_R}$. Let us ignore pulse broadening in time due to dispersion. The maximum number of orthogonal temporal modes that can be packed within the probing interval $T$ is roughly equal to $M_T = WT$. 

Using the Yuen-Shapiro quantum diffraction theory \cite{yuen1978optical}, the field at the receiver $\hat{E}_L(\mb{\rho_L}, t) :=\hat{E}(\mb{r},t)|_{z=L}$ is connected to the field at the transmitter $\hat{E}_0(\mb{\rho_0},t) :=\hat{E}(\mb{r},t)|_{z=0}$ via the Huygens-Fresnel diffraction integral: $\hat{E}(\mb{\rho}_L,t) = \int d^2\mb{\rho}_0 \hat{E}(\mb{\rho}_0,t-L/c)h(\mb{\rho}_0-\mb{\rho}_L)$. Here $h(\mb{\rho}) = \exp\left[ikL + ik|\mb{\rho}|^2/2L\right]/(i\lambda L)$, is a linear space-varying impulse response \cite{yuen1978optical}, which admits a normal-mode decomposition,
$
h(\mb{\rho}_0-\mb{\rho}_L) = \sum_n \sqrt{\eta_n}\Phi_n(\mb{\rho}_L)\phi_n(\mb{\rho}_0)
$
where $k = 2\pi/\lambda$ is the wavenumber and $\brac{\eta_n}$ are arranged s.t. $0<\eta_0<\eta_1<\ldots <1$. Here $\brac{\phi_n(\mb{\rho}_0)}$ and $\brac{\Phi_n(\mb{\rho}_L)}$ are the {\em normal modes}, complete orthonormal sets of modes at the transmitter and receiver planes, respectively, such that if only the $\brac{\phi_n(\mb{\rho}_0)}$ mode is modulated at the transmitter aperture, only the $\brac{\Phi_n(\mb{\rho}_L)}$ mode will be excited at the receiver aperture, but with amplitude attenuation $\brac{\eta_n}$. 

Physically, this decomposition implies that diffraction-limited propagation of a general optical quantum field between two apertures can be thought of as a countably-infinite set of independent lossy bosonic channels:
$
\hat{a}_n^{(L)} = \sqrt{\eta_n}\hat{a}_n^{(0)}+\sqrt{1-\eta_n}\hat{e}_n~,
$
where $\hat{\mb{a}}_{0} := (\hat{a}_0^{(0)},\hat{a}_1^{(0)},\ldots)$ and $\hat{\mb{a}}_{(L)} := (\hat{a}_0^{(L)},\hat{a}_1^{(L)},\ldots)$  are the annihilation operators corresponding to the transmitter and receiver pupil normal modes, respectively. $\brac{\hat{e}_n}$ are the annihilation operators of environment modes we must include to preserve commutator brackets. In the near-field regime, i.e., Fresnel number product $D = (\pi r_Tr_R/\lambda L)^2\gg 1$, there are roughly $D$ modes that are essentially lossless, i.e., $\eta_n \approx 1$, for $0 \leq n < D$~\cite{yuen1978optical}.

Now consider a beam displacement $\mb{d} = (d_x, d_y)$ or a rotation $\theta = \abs{\mb{d}}/L$ of the transmitted field. As long as the displacement is small compared to the size of the receiver's aperture, i.e., $\abs{\mb{d}}/r_R\ll 1$, these two scenarios can be considered as equivalent. Since the measurement is applied on the received field, we consider the equivalent situation in which the receiver's aperture is displaced by $-\mb{d}$. Assuming the receiver separates the vacuum-propagation normal modes $ \brac{\Phi_n(\mb{\rho}_L)}$ (since it does not know $d$ apriori), the multi-spatial-mode input-output relationship is no longer an array of independent beamsplitters. The displacement induces modal cross talk, which can be seen as a spatial-mode transformation,

$\label{eq:unitary}
\hat{\mb{a}}_L \rightarrow U(d)\hat{\mb{a}}_L U(d)^\dagger = \mb{S} \hat{\mb{a}}_L
$.
We can see that the action of displacement is a passive Gaussian unitary transformation~\cite{weedbrook2012gaussian}. The {\em coupling matrix} $\mb{S}$ is given by the following overlap integrals between the original and the displaced receiver-pupil normal modes:
\begin{align}\label{eq:coupling-matrix}
\mb{S}_{mn}(\mb{d}) &= \int d^2\mb{\rho}_L \Phi_m^*(\mb{\rho}_L-\mb{d})\Phi_n(\mb{\rho}_L)~.
\end{align}
Therefore, the action of the beam displacement on a general multi-spatial-mode quantum state is the unitary $U(\mb{d}) = \exp\sbrac{-\hat{\mb{a}}^\dagger_L(\ln \mb{S}(d))\hat{\mb{a}}_L}$. We should note here that the transformation is unitary since we are assuming the transmitter to be only modulating the lossless modes. If the transmitter modulates more than $D$ modes, or just one spatial mode in the far field regime ($D < 1$), we must take the losses ($\eta_n$) into account.

Several simplifications are in order. First, in this work we will restrict ourselves to a single-scalar-parameter estimation problem, by assuming that the direction of displacement (in the $(x,y)$ plane) is known to the receiver a priori. Without loss of generality, we choose that direction to be the $x$-axis, i.e., $\mb{d}=(d_x, 0)$. Secondly, in the regime of the displacement being small, i.e., $\tilde{d} := d_x/r_R\ll 1$, we will just keep up to the leading order term in $\tilde{d}$ in the coupling matrix
$
\mb{S}= \mb{I}- \mb{\Gamma}\tilde{d}+ O(\tilde{d}^2),
$
where,
\begin{align}
\mb{\Gamma}_{mn} = r_R\int_{-\infty}^\infty dxdy\frac{\partial \phi_m^*(x,y)}{\partial x}\phi_n(x,y)~.
\end{align}
It is evident that $\mb{\Gamma}$ is anti-Hermitian, i.e, $\Gamma_{mn}=-\Gamma^*_{nm}$. The unitary in this limit is given by $U(\mb{d}) = \exp(i\tilde{d}\hat{H})$, where
\begin{align}\label{eq:Hamiltonian}
\hat{H} =i \hat{\mb{a}}^\dagger_L{\mb{\Gamma}}\hat{\mb{a}}_L~.
\end{align}

The Fresnel number product $D$ separates all normal modes roughly into two sets: lossless and lossy modes. In our 1D problem, fixing the mode index along the $y$ direction to zero, the number of lossless spatial modes available to us is roughly $M_S :=\sqrt{D}$. Therefore, we will only modulate the first $M_S$ modes, since loss is known to be detriment to quantum enhancements in metrology \cite{demkowicz2012elusive}. At first glance, the mode-coupling matrix in Eq.~\eqref{eq:coupling-matrix} induced by the beam displacement seems to make this truncation impossible. However, intuitively, the spatial mode cross talk should be ``short-ranged" (e.g., nearest neighbor in the mode indices) for infinitesimal displacements. As long as we discard all the modes with indices above $M_S-\kappa$, where we define the maximal coupling range $\kappa = \min\brac{k: \mb{\Gamma}_{m, m+\kappa+1} = 0}$, the leftover subset of modes stays lossless. 

For circular hard apertures, the normal modes are the generalized prolate-spheroidal wavefunctions, the analytical form of which are involved~\cite{slepian1964prolate,slepian1965analytic}. To clearly illustrate the truncation procedure, we will assume Gaussian-attenuation aperture pupils whose normal modes are Hermite-Gaussian (HG) modes~\cite{shapiro2005ultimate}, 
$
\Phi_n(x) = \pbrac{\frac{2}{r_R^2}}^{\frac{1}{4}}\psi_n\pbrac{\frac{\sqrt{2}x}{r_R}}
$.
Here $\psi_n(x)=\ (2^nn!\sqrt{\pi})^{-\frac{1}{2}}e^{-x^2/2}H_n(x)$ is the Hermite polynomial. We simply ignore the phase factor, since it does not contribute to $\mb{\Gamma}$ and our unitary.
For HG modes we have $\kappa = 1$, that is, only  nearest-neighbor couplings exist, as can be seen by directly calculating the coupling matrix \cite{supp},
\begin{align}\label{eq: Gamma}
\mb{\Gamma}_{mn}={\sqrt{m}\delta_{m-1,n}-\sqrt{m+1}\delta_{m+1,n}}~.
\end{align}
Therefore, the first $M_S-1$ modes comprise a closed lossless subspace under the action of {\em small} beam displacements. 

\begin{figure}[htp]
	\centering
	\includegraphics[width=0.9\columnwidth]{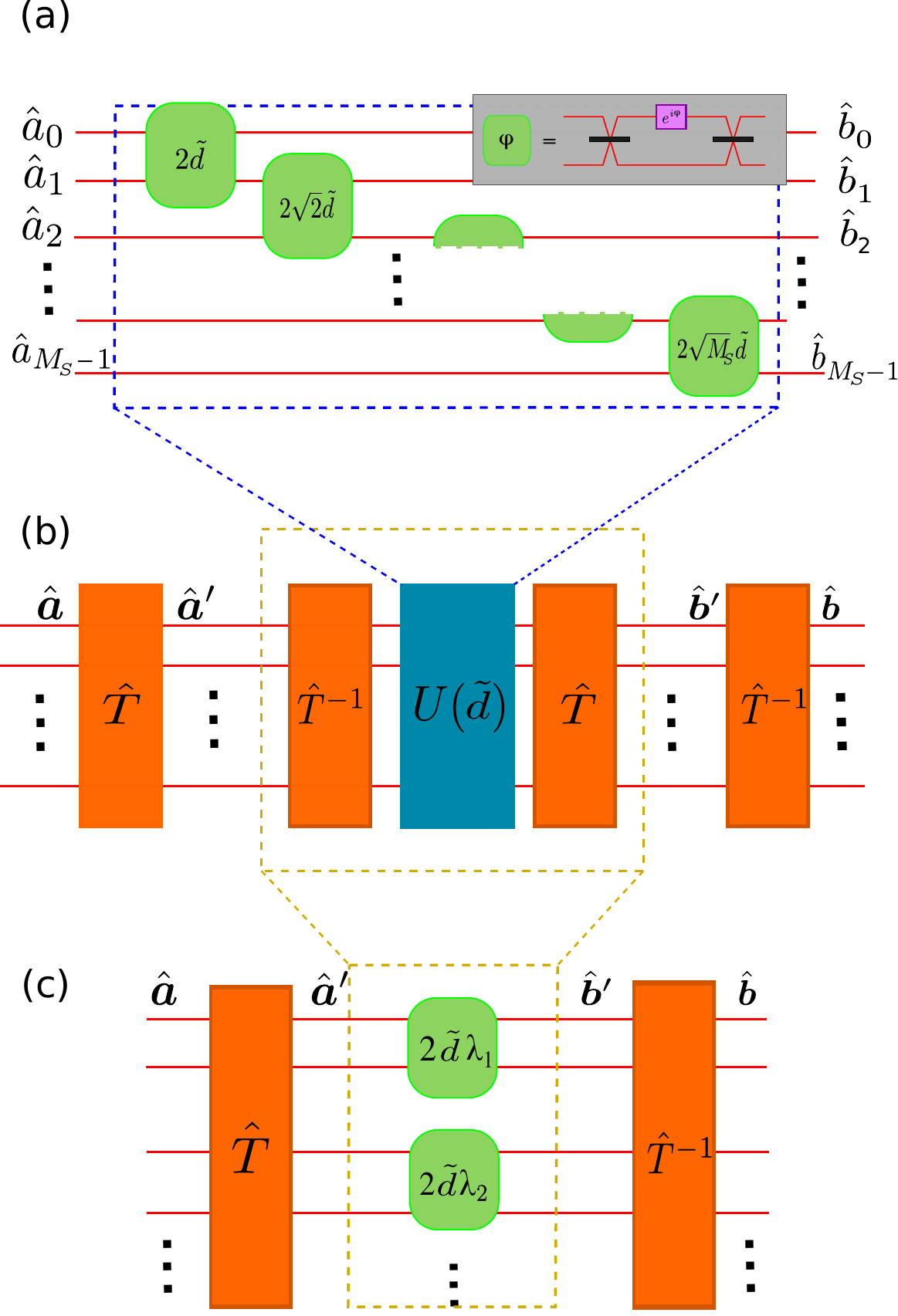}
	\caption{(a) Unitary quantum model of beam displacement $\tilde{d}$. In the limit of $\tilde{d}=d/r_R \ll 1$ where $r_R$ is the radius of the receiver aperture, and the near field regime ($D \gg 1$), the effect of beam displacement is a series of pairwise nested Mach-Zehnder interferometer (MZI) interactions on spatial modes $n$ and $n-1$, $n = 1, \ldots, M_S - 1$. The $n$-th MZI consists of a phase shift of $2\sqrt{n}\tilde{d}$ sandwiched by two 50-50 beam-splitters. (b) By inserting a properly chosen mode transformation $\hat{T}$ and its inverse on either side of $U(\tilde d)$, we can show that the effective beam displacement unitary ${\hat T}^{-1}U(\tilde d){\hat T}$ in the transformed mode basis is a set of $M_S/2$ pairwise two-mode MZIs, as shown in (c). The phase of each MZI is given by the eigenvalues of the coupling matrix $\mb{\Gamma}$ described in the text, multiplied by $\tilde d$. \label{fig:unitary}}
\end{figure}

In summary, our quantum model is fully described by the unitary 
$
U(\tilde{d})=\exp\sbrac{i\tilde{d}\hat{H}},
$
where 
$
\hat{H}=i\sum_{n=1}^{M-1}\sqrt{n}\sbrac{\hat{a}_{n}^\dagger\hat{a}_{n-1}-\hat{a}^\dagger_{n-1}\hat{a}_{n} }
$
by using Eq.~\eqref{eq: Gamma}. Hereafter we will not differentiate between the mode operators at the transmitter and those at receiver, since they are the same for the first $M_S$ modes. 

Using the Jordan-Schwinger map \cite{sakurai1995modern}, $\hat{J}^n_x=\frac{1}{2}(\hat{a}^\dagger_{n-1} \hat{a}_{n} +\hat{a}^\dagger_{n} \hat{a}_{n-1}), \hat{J}^n_y=\frac{i}{2}(\hat{a}_{n}^\dagger\hat{a}_{n-1}-\hat{a}^\dagger_{n-1}\hat{a}_{n}), \hat{J}^n_z=\frac{1}{2}(\hat{a}^\dagger_{n-1} \hat{a}_{n-1} -\hat{a}^\dagger_{n} \hat{a}_{n})$, the Hamiltonian can be compactly  written as follows
\begin{align}\label{eq: Hamiltonian-HG}
\hat{H} = \sum_{n=1}^{M_S-1} 2\sqrt{n}\hat{J}_y^n~.
\end{align}
Each term in the above sum represents a MZI with phase $2\sqrt{n}\tilde{d}$ \cite{demkowicz2015quantum}. Therefore, in the limit $\tilde{d} \ll 1$, the unitary operator $U({\tilde d})$ that captures the effect of a small beam displacement factorizes into a form where each mode interacts with its two neighboring modes with  a two-mode MZI, as shown in Fig.~\ref{fig:unitary}.

In the next three sections, we will quantify the performance of the sensor using the quantum Cram\'{e}r-Rao bound, which is given by the inverse of QFI. Given $\nu$ copies of the state $\rho_d$ (which encodes parameter $d$), it gives a lower bound on the variance of an unbiased estimator constructed from joint quantum measurement at the output, i.e,
\begin{align}\label{eq:QCRB}
\delta d^2 \geq \frac{1}{\nu\mathcal{F}_Q(\rho_d)}.
\end{align}

Quantum Cram\'{e}r-Rao bound is a tighter lower bound compared with that given by the classical Cram\'{e}r-Rao bound \cite{sengijpta1995fundamentals} of the outcome of any specific quantum measurement on $\rho_d$ (see Appendix~\ref{app:FisherInfo}). For a unitary of the form $U = \exp[i\tilde{d}\hat{H}]$, ${\tilde d} = d/r_R$ and a pure input state probe, resulting in an output $\rho_d$, the QFI is independent of $d$ and is simply a constant times the variance $\Delta^2\hat{H} = {\ave{\hat{H}^2}-\ave{\hat{H}}^2}$ of the Hamiltonian $\hat{H}$, i.e.,
$
\F_Q = ({4}/{r_R^2})\Delta^2\hat{H}.
$

\section{Optimum classical probe}\label{sec:classical}
We first derive the minimum estimation error that could be achieved by a general single-spatial-mode probe state (classical or quantum). Let us consider a probe whose $j$th spatial-mode is excited in some state $
|\psi\rangle$ with mean photon number $N_S$, $j\in[0,M_S-2]$, while leaving the other spatial modes in vacuum. The calculation of the variance of the Hamiltonian $\hat H$ is straightforward. The mean value vanishes due to the skew-symmetry of the coupling matrix $\mb{\Gamma}$, i.e., $\mb{\Gamma}^T = -\mb{\Gamma}$. 
$\ave{\hat{H}} = i\ave{\mb{\Gamma}_{00}\hat{a}^\dagger_0\hat{a}_0} = 0$. For the mean square of $\hat{H}$, only the coupling term between the $j$ and $j+1$ modes contributes. Therefore, we have $\Delta^2\hat{H} =N_S\mb{\Gamma}_{j,j+1}^2 = jN_S$,
which gives $\F_Q = 4jN_S/r_R^2$. From Eq.~\eqref{eq:QCRB} in Appendix~\ref{app:FisherInfo} we know that the minimum error that can be achieved by a single-mode state is $\delta d = r_R/2\sqrt{jN_S}$.

Therefore, we conclude that, a single-spatial-mode probe: (1) cannot surpass the SQL, i.e., $1/\sqrt{N_S}$ scaling, but (2) populating a higher-order spatial mode (i.e., higher mode index $j$) achieves a better sensitivity.

At first glance, conclusion (1) is rather surprising, considering the fact that the output of the effective multi-mode interferometer in Fig.~\ref{fig:unitary} even if only one of the input modes is excited (e.g., in a squeezed state) with the other inputs in vacuum, is in general an entangled state. But, (1) is consistent with the finding in Ref.~\cite{fabre_quantum_2000}. The analysis in~\cite{fabre_quantum_2000} leading to their conclusion (that a single-spatial-mode probe cannot beat SQL scaling), however, was restricted to the case of a split-detector receiver. Their result was not conclusive since the most general receiver measurement was not accounted for. Our QFI-based result conclusively rules out the possibility of surpassing SQL scaling with a single-spatial-mode probe, and provides the impetus to consider multi-spatial-mode (and multi-spatio-temporal mode) entangled states.

Our conclusion (2) above, that populating a higher-order mode is able to achieve better accuracy, can be intuitively understood by noticing that a higher-order HG mode oscillates (in space) more rapidly, thereby making it more sensitive to a small transverse spatial shift of the beam. Mathematically, this shows up as the $\sqrt{j}$ pre-factor in the effective MZI phase accrued in the interference between modes $j-1$ and $j$, as illustrated in Fig.~\ref{fig:unitary}. In other words, physically, probing with a high-order spatial mode once is equivalent to probing with a lower-order mode multiple times, since the same beam displacement results in the higher-order mode getting modulated by a larger phase. We should emphasize here that this result is not restricted to HG modes, but is true for any choice of aperture function (and associated normal modes) \cite{supp}.


Now we are ready to derive the performance of the optimal classical probe. The most general $M_S - 1$ mode classical state is a mixture of product of coherent states $\int d\mb{\alpha}P(\mb{\alpha})\ket{\mb{\alpha}}\bra{\mb{\alpha}}$, where $\mb{\alpha}:=\pbrac{\alpha_0,\ldots, \alpha_{M_S-2}}$ and $P(\mb{\alpha})$ is arbitrary probability distribution. As previously mentioned, it is sufficient to consider a pure input state thanks to the convexity of the QFI~\cite{fujiwara2005quantum}. So, considering a coherent state $|{\mb{\alpha}}\rangle$ suffices. The next crucial observation is that a coherent state is {\em always} single (spatio-temporal) mode in an appropriate mode basis
~\footnote{In order to see why, we observe that a linear-optical unitary acting on a product coherent state (expressed in some orthonormal mode basis),
	{$\ket{\mb{\alpha}}$ } produces another product coherent state (in that same mode basis) ${|{\mb{\beta}}\rangle}$ with $\mb{\beta} = {\mb{U}} \mb{\alpha}$, with ${\mb{U}}$ a complex-valued unitary matrix. Therefore the coherent state ${|{\mb{\alpha}}\rangle}$ can always be thought of as a single-mode coherent state in an appropriate mode basis, and thus a unit vector in an orthonormal set constructed via a Gram-Schmidt orthogonalization where rest of the modes are in their vacuum states.}.

Now invoking our above result for the general single-mode quantum state, the optimal precision is obtained by putting the coherent state in the highest-order normal mode, yielding:
\begin{align}\label{eq:classical-error-spatial}
\delta d^P\simeq \frac{r_R}{2\sqrt{M_SN_S}} = \frac{r_R}{2M_S\sqrt{\bar n}}~,
\end{align}
where ${\bar n} = N_S/M_S$ is the mean photon number per mode (ignoring the difference between $M_S$ and $M_S-1$). 

To generalize the above result to spatio-temporal modes, considering a product of $M_T$ single-spatial-mode states with precision given in Eq.~\eqref{eq:classical-error-spatial}, given the QFI is additive, we have 
\begin{align}\label{eq:classical-error}
\delta d^P\simeq \frac{r_R}{2\sqrt{M_SM_TN_S}}= \frac{r_R}{2M_S\sqrt{M_T\bar{n}}}~.
\end{align}
Eq.~\eqref{eq:classical-error} also follows readily from~\eqref{eq:classical-error-spatial} by replacing $\bar n$ with $M_T{\bar n}$; the rationale being, a coherent state is always single mode, i.e., we can reinterpret the optimal probe as a single spatio-temporal mode coherent state with $M_T{\bar n}$ mean photons in the highest-order normal mode.

\section{Optimum entangled probe}\label{sec:multimode}
We first show that the Hamiltonian in Eq.~\eqref{eq: Hamiltonian-HG}, which describes a set of coupled MZIs, can be transformed into one of a set of independent MZIs (as we will show in Eq.~\eqref{eq:Hamiltonian-MZIs}), after a suitable unitary mode-transformation. The problem of finding the optimal multi-mode probe state thereby reduces to finding the optimal probe in a new mode basis, where each mode pair accrues an independent phase (see Fig.~\ref{fig:unitary}). Again, we start by focusing on spatial modes, i.e., fixing a particular temporal mode index, and then generalize to the case of using full spatio-temporal modes at the end of this section. 

We first insert two pairs of unitaries $\brac{\hat{T},\hat{T}^\dagger}$ without changing the dynamics, as shown in Fig.~\ref{fig:unitary} (b), i.e.,
\begin{align}\label{eq:tranformation-a}
\hat{\mb{a}}':=\hat{T}\hat{\mb{a}}\hat{T}^{\dagger} \equiv \mb{T}\hat{\mb{a}}~,
\end{align}
where $\mb{T}$ is the transformation matrix on the annihilation operators induced by the unitary $\hat{T}$. 

For a skew-symmetric matrix $\mb{\Gamma}$ (i.e., $\mb{\Gamma}^T = -\mb{\Gamma}$), there exists an orthogonal transformation $\mb{T}$ \cite{eves1966elementary}, such that
\begin{align}\label{eq:diagonalization}
\mb{T}\mb{\Gamma}\mb{T}^T = \bigoplus_{k=1}^{\lceil M_S/2 \rceil} i\mb{\sigma}_y\lambda_k~,
\end{align}
where $\mb{\sigma}_y = \left(\begin{array}{cc} 0 & -i \\ i & 0 \end{array}\right)$ is the Pauli Y operator, and $\brac{\pm i\lambda_k}$ are the eigenvalues of the coupling matrix $\mb{\Gamma}$~\footnote{For $M_S$ odd, the last row and column of the transformed $\mb{\Gamma}$ matrix are zeros and thus can be dropped. Therefore, we can take $M_S$ to be even without loss of generality}. In general, finding $\brac{\lambda_k}$ requires solving the roots of the characteristic equation of $\mb{\Gamma}$, for which no analytical formula exists. 

We choose $\hat{T}$ that brings $\mb{\Gamma}$ into it's aforesaid `normal form'~\eqref{eq:diagonalization}. The fact that $\mb{T}$ is orthogonal implies that $\hat{T}$ is a passive Gaussian unitary~\cite{weedbrook2012gaussian}, and hence realizable by a mode transformation. 

To re-express the Hamiltonian of Eq.~\eqref{eq:Hamiltonian} in the new basis $\hat{\mb{a}}^{\prime}$, we apply Eq.~\eqref{eq:tranformation-a} and have   $\hat{H} = i\hat{\mb{a}}^{\prime\dagger}\pbrac{\mb{T}\mb{\Gamma}\mb{T}^T}\hat{\mb{a}}^{\prime}$. Invoking the transformation in Eq.~\eqref{eq:diagonalization}, we have
\begin{align}\label{eq:Hamiltonian-MZIs}
{\hat H} = 2\sum_{k=1}^{\lceil M_S/2\rceil} \lambda _k \hat{S}^{2k-1}_y~,
\end{align}
where
$\hat{S}^{2k-1}_y =\frac{i}{2}(\hat{a}_{2k-1}^{\prime\dagger}\hat{a}^\prime_{2k-2}-\hat{a}^{\prime\dagger}_{2k-2}\hat{a}_{2k-1}^\prime)$. Since each term in the above sum describes an MZI with phase $2\tilde{d}\lambda_k$ \cite{demkowicz2015quantum}, we have re-expressed the action of beam displacement---originally expressed in Eq.~\eqref{eq: Hamiltonian-HG} as a nearest-neighbor-mode coupled unitary on the $\hat{\mb{a}}$ modes---to a pairwise-mode coupled unitary where pairs of $\hat{\mb{a}}'$ modes accrue independent MZI phases (See Fig.~\ref{fig:unitary} (c)), as described by Eq.~\eqref{eq:Hamiltonian-MZIs}.  For later convenience, we define $N_k$ to be the average photon number put into the $k$th subsystem, i.e., in modes $2k-1$ and $2k-2$.

To construct the $M_S$-mode (entangled) state which maximizes the QFI, $\F_Q = ({4}/{r_R^2})\Delta^2\hat{H}^\prime$, we first consider an upper bound  $\Delta\hat{H}^\prime\leq \sum_k (s_{\max}^k-s_{\min}^k)/2$ \cite{giovannetti2006quantum}, where $s_{\max}^k(s_{\min}^k)$ is the maximum (minimum) eigenvalue of the $k$th two-mode subsystem described by Hamiltonian $2\lambda _k \hat{S}^{2k-1}_y$. From the Schwinger representation \cite{schwinger1965quantum, yurke19862}, each subsystem with Hamiltonian $2\lambda_k \hat{S}^{2k-1}_y$ is equivalent to a spin-$N_k/2$ system, thereby we have $s_{\max}^k =\lambda_kN_k$ and $s_{\min}^k =-\lambda_kN_k$. Summing them together we have $\Delta^2\hat{H}' \leq (\sum_k\lambda_k N_k)^2$.

The optimal probe that saturates this upper bound is readily given by the following entangled state in the $\mb{a}'$ basis~\cite{giovannetti2006quantum}:
\begin{align}
\ket{\Psi^E}_{\mb{a}'} &= \frac{1}{\sqrt{2}}\pbrac{\ket{+}_{\mb{a}'}+\ket{-}_{\mb{a}'}}~, \,{\text{with}}\\
\ket{+}_{\mb{a}'}&=\bigotimes_{k=1}^{[M_S/2]} \hat{S}^{2k-1}_x\pbrac{\frac{\pi}{2}} \ket{N_k,0}_{a'_{2k-1}a'_{2k}}~,\\
\ket{-}_{\mb{a}'}&=\bigotimes_{k=1}^{[M_S/2]} \hat{S}^{2k-1}_x\pbrac{\frac{\pi}{2}} \ket{0,N_k}_{a'_{2k-1}a'_{2k}}~.
\end{align}
The states $\ket{\pm}_{\mb{a}'}$ correspond to putting all the spins into up (resp., down) along the $y$ direction. Here $\hat{R}_x^{2k-1}=\frac{1}{2}(a^{\prime\dagger}_{2k-1}a_{2k}^\prime+a^{\prime\dagger}_{2k}a_{2k-1}^\prime)$. The optimal probe in the original $\hat{\mb{a}}$ mode basis is readily obtained by applying the $M_S$-mode linear transformation $\hat{T}^\dagger$ on $\ket{\Psi^E}_{\mb{a}'}$. 

For a given photon-number distribution across spatial modes $\brac{N_k}$, the optimal QFI achieved by this entangled probe is given by $4(\sum_k\lambda_k N_k)^2/r_R^2$. However, we can further optimize the QFI over all possible photon number distributions. Define ratio $c_k = N_k/N_S$ such that $\sum_{k=1}^{\lceil M_S/2\rceil}c_k=1$. The QFI given by $4(\sum_k\lambda_k N_k)^2/r_R^2 =4N_S^2(\sum_k\lambda_k c_k)^2/r_R^2 $ is maximized by choosing $c_k =\lambda_k/\sum_k\lambda_k$.
Finally, we have the optimal QFI achieved by this choice of photon distribution, 
\begin{align}\label{eq:entangled-QFI-norm}
\F^E_Q = \frac{4N_S^2}{r_R^2}\pbrac{\frac{\sum_k \lambda_k^2}{\sum_k\lambda_k}}^2~.
\end{align}

To study the asymptotical behavior of the QFI, notice that $\sum_k\lambda_k^p = \frac{1}{2}\Vert \mb{\Gamma}\Vert_p^p$, 
where $\Vert \mb{\Gamma}\Vert_p$ is the Schatten $p$-norm of $\mb{\Gamma}$. In the limit of $M_S\gg 1$ we have $\Vert\mb{\Gamma}\Vert_2\simeq M_S$ and $\Vert\mb{\Gamma}\Vert_1\simeq M_S^{3/2}$ \cite{supp}.
Therefore, we have following minimum estimation error,
\begin{align}\label{eq:entangled-rms-spatial}
\delta d^E \simeq \frac{r_R}{\sqrt{M_S}N_S}~.
\end{align}
This super-Heisenberg scaling behavior is actually a composite effect of the spatial entanglement of the probe \cite{giovannetti2006quantum} and the phase gradient in our Hamiltonian $\hat H$ in~\eqref{eq:Hamiltonian-MZIs}~\cite{d1997arbitrary}, i.e., increasing $\lambda_k$ values in the effective MZI array in Fig.~\ref{fig:unitary}(c). The former contributes to the HL scaling while the latter leads to another $\sqrt{M_S}$ enhancement in the sensitivity.

So far we have been considering spatial modes. Our results can be readily generalized to include the use of all temporal modes available. If we don't entangle across the temporal mode index, i.e., consider a product state over the $M_T$ orthogonal temporal modes, we have the following precision for this hybrid probe (entangled in space but not over time), we get
\begin{align}
\delta_d^H \simeq \frac{d_R}{\sqrt{M_T}M_S^{3/2}\bar{n}}~,
\end{align}
from the additivity of the QFI.

On the other hand, the optimal spatio-temporal probe state is an entangled state across {\em both} the spatial and the temporal indexes. For $M_T$ temporal-modes, effectively we have $M_T$ copies of the original coupling matrix, $\bigoplus_{i=1}^{M_T}\mb{\Gamma}$. Therefore, by simply redefining $c_k =\lambda_k/ \sum_{k=1}^{M_SM_T/2}\lambda_k$, we get an optimal QFI with the same form as in Eq.~\eqref{eq:entangled-QFI-norm}, with $N_S$ replaced by $N$. It is not difficult to see that the terms inside of the bracket in Eq.~\eqref{eq:entangled-QFI-norm} stay the same, thanks to the periodicity of $\brac{\lambda_k}$. Therefore, the best precision obtained by using a probe entangled across all the spatio-temporal modes is given by
\begin{align}\label{eq:entangled-rms}
\delta d^E\simeq \frac{r_R}{\sqrt{M_S}N}=\frac{r_R}{M_TM_S^{3/2}\bar{n}}~.
\end{align}

\section{Structured squeezed-light transceiver}\label{sec:structured}
Although we found the optimal spatio-temporally-entangled quantum probe in section~\ref{sec:multimode}, designing an explicit transmitter and receiver design for that probe is difficult. In this section, we construct a fully structured transceiver design that involves a Gaussian (multi spatio-temporally-entangled squeezed-state) probe and a Gaussian (homodyne-like) measurement, which attains the quantum-optimal scaling with respect to $M_T$, $M_S$ and $\bar n$, as in Eq.~\eqref{eq:entangled-rms}. We again first consider the $M_T=1$ case, since generalization to $M_T>1$ is straightforward. 

\begin{figure}[t]
	\centering
	\includegraphics[width=\columnwidth]{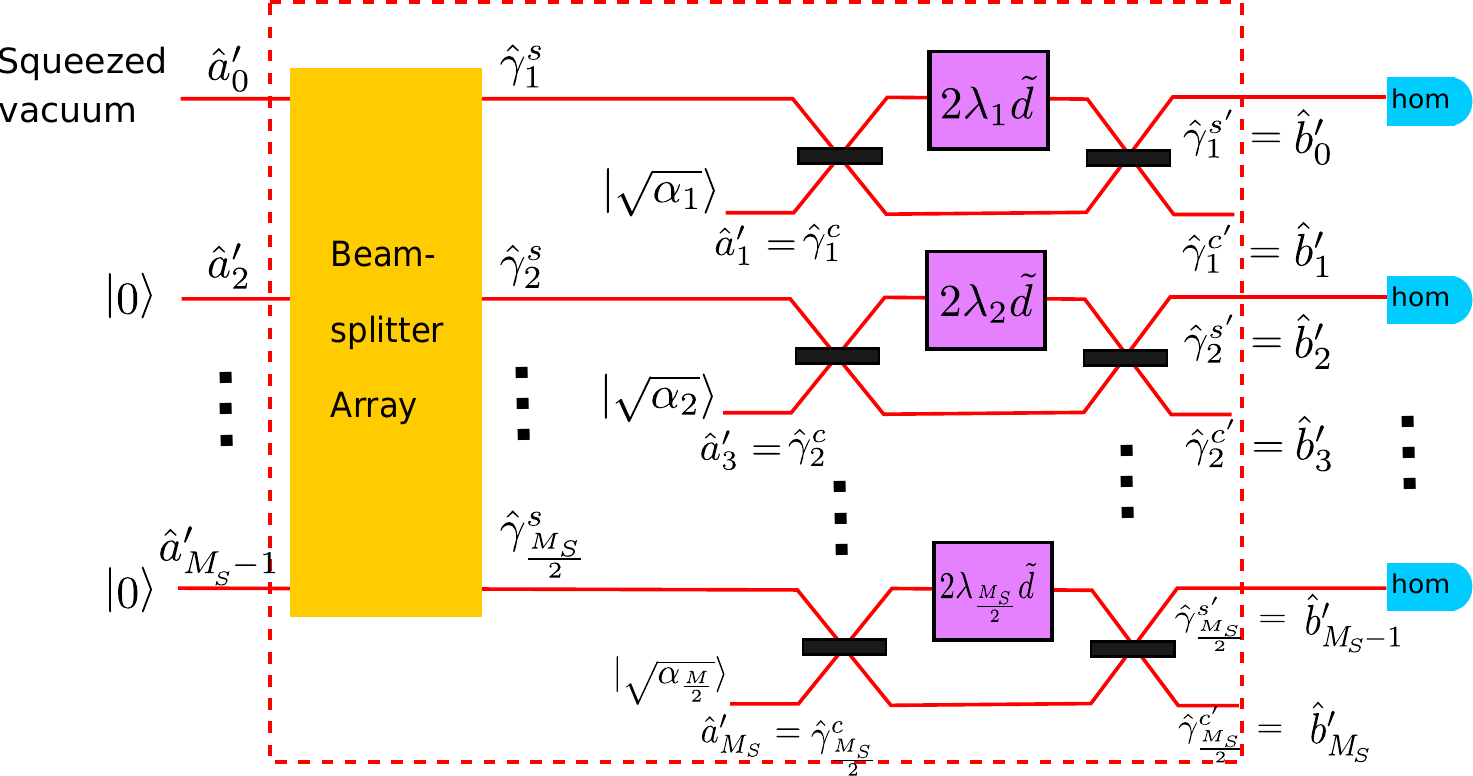}
	\caption{A schematic of the Gaussian multi-mode-entangled transmitter-receiver pair that achieves optimal scaling of the sensitivity of estimating beam displacement. The $M_S/2$ MZIs shown are those in Fig.~\ref{fig:unitary}(c), and the state input into those are of the ${\hat {\bm a}}^\prime$ modes. The actual probe (the ${\hat {\bm a}}$ modes) are related to the ${\hat {\bm a}}^\prime$ modes by a mode transformation $\hat T$. The beam displacement $\tilde{d}$ gets encoded as a quadrature displacement of the probe light. The receiver is an array of homodyne receivers on the ${\hat {\bm b}}^\prime$ modes (see Fig.~\ref{fig:unitary}(c)). Since the target-modulated state, the ${\hat {\bm b}}$ modes, are a mode transformation ${\hat T}^{-1}$ away from the ${\hat {\bm b}}^\prime$ modes, the actual sensor receiver must be an appropriately mode-resolved homodyne array. See Fig.~\ref{fig:fullGaussian}.}
	\label{fig:GaussianCircuit}
\end{figure}
To construct the Gaussian state that achieves the scaling in Eq.~\eqref{eq:entangled-rms}, we consider the setup shown in Fig.~\ref{fig:GaussianCircuit}. The mode pairs that interrogate the $M_S/2$ decoupled effective MZIs (see Fig.~\ref{fig:unitary}(b)) are $\hat{\mb{\gamma}}:=\brac{\hat{\mb{\gamma}}^s, \hat{\mb{\gamma}}^c}$. The mean transmit photon number across all spatial modes, $N_S$, is distributed equally between the $\hat{\mb{\gamma}}^s$ and $\hat{\mb{\gamma}}^c$ modes, i.e., $N_s = N_c = N_S/2$. The modes ${\hat{\mb{\gamma}}^s} \equiv (\hat{\gamma}_1^s, \ldots, \hat{\gamma}_{M_S/2}^s)$ are a result of a linear mode transformation (a beamsplitter array, to be explicitly defined later) applied on the even $\hat{\mb{a}}'$ modes. The modes ${\hat{\mb{\gamma}}^c} \equiv (\hat{\gamma}_1^c, \ldots, \hat{\gamma}_{M_S/2}^c)$ are excited in coherent states $\ket{\sqrt{\alpha_k}}$, $k = 1, \ldots, M_S/2$ with mean photon number commensurate with the phase gradient in the effective MZI array, i.e., $|\alpha_k|^2 = c_kN_c$, $c_k =\lambda_k/\sum_k\lambda_k$, with $\lambda_k$ as in section~\ref{sec:multimode}. 

In near-field applications where the number of lossless spatial modes $N_S$ is large, and the beam displacement to be measured is small, we have $\lambda_k\tilde{d}\ll 1$. In this regime, the output modes from the MZI array can be approximated as~\cite{zhuang2018distributed}
$
\hat{\gamma}^{s'}_k \simeq (1-i{\lambda_k\tilde{d}})\hat{\gamma}^s_k +i{\lambda_k \tilde{d}}\hat{\gamma}_k^c~.
$
We see that the beam displacement $\tilde d$ gets encoded into mean fields (quadrature displacements) of the originally-zero-mean $\hat{\gamma}_k^s$ modes. In other words, $\langle \hat{\gamma}_k^{s^\prime} \rangle = \lambda_k\ave{\hat{\gamma}_k^c}\tilde{d} = \lambda_k\sqrt{c_kN_c}\tilde{d}$. Consider the following estimator of $\tilde{d}$:
\begin{align}\label{eq:estimator}
\hat{d} &= \frac{\sqrt{2}}{A\sqrt{N_S}}\sum_{k=1}^{\lceil M_S/2\rceil} \sqrt{c_k}~ \text{Im}\pbrac{\hat{\gamma}_k^{s'}}~,
\end{align}
where $A= {\sum_k \lambda_k^2}/{\sum_k\lambda_k}$. It is straightforward to check that the estimator constructed above is unbiased, in the sense that $\ave{\hat{d}}=\tilde{d}$. 

Now we choose the beamsplitter array in Fig.~\ref{fig:GaussianCircuit} to be a unitary such that
$
\hat{a}_0' =\sum_k \sqrt{c_k}\hat{\gamma}_k^s,
$
which is possible since it preserves the canonical relation $\sbrac{\hat{a}'_0, \hat{a'_0}^{\dagger}}=1$. We thus have
$
\hat{d} = \frac{\sqrt{2}}{A\sqrt{N}}\text{Im}\pbrac{\hat{b}'_0}.
$
The estimation error is minimized if the $\hat{a}'_0$ mode is in its squeezed-vacuum state with an average photon number $N_s$ whose real quadrature is squeezed \cite{yuen1978optical,zhuang2018distributed},
\begin{align}\label{eq:Gaussian-error-spatial}
\delta\tilde{d}^G &= \frac{\sqrt{2}}{A\sqrt{N_S}}\frac{1}{\sqrt{N_s+1}+\sqrt{N}_s}~.
\end{align}

Taking the large $N_S$ limit, and using the facts $N_s=N_S/2$, $A\simeq \sqrt{M_S}$ and $\tilde{d}=d/r_R$, the Gaussian state we constructed above achieves the same estimation error scaling as in Eq.~\eqref{eq:entangled-rms}. Notice that if we set $N_s=0$ in Eq.~\eqref{eq:Gaussian-error-spatial}, we have a classical input state, and the estimation error scaling is consistent with our previous result for the optimal classical probe in Eq.~\eqref{eq:classical-error}.

\begin{figure}[t]
	\centering
	\includegraphics[width=\columnwidth]{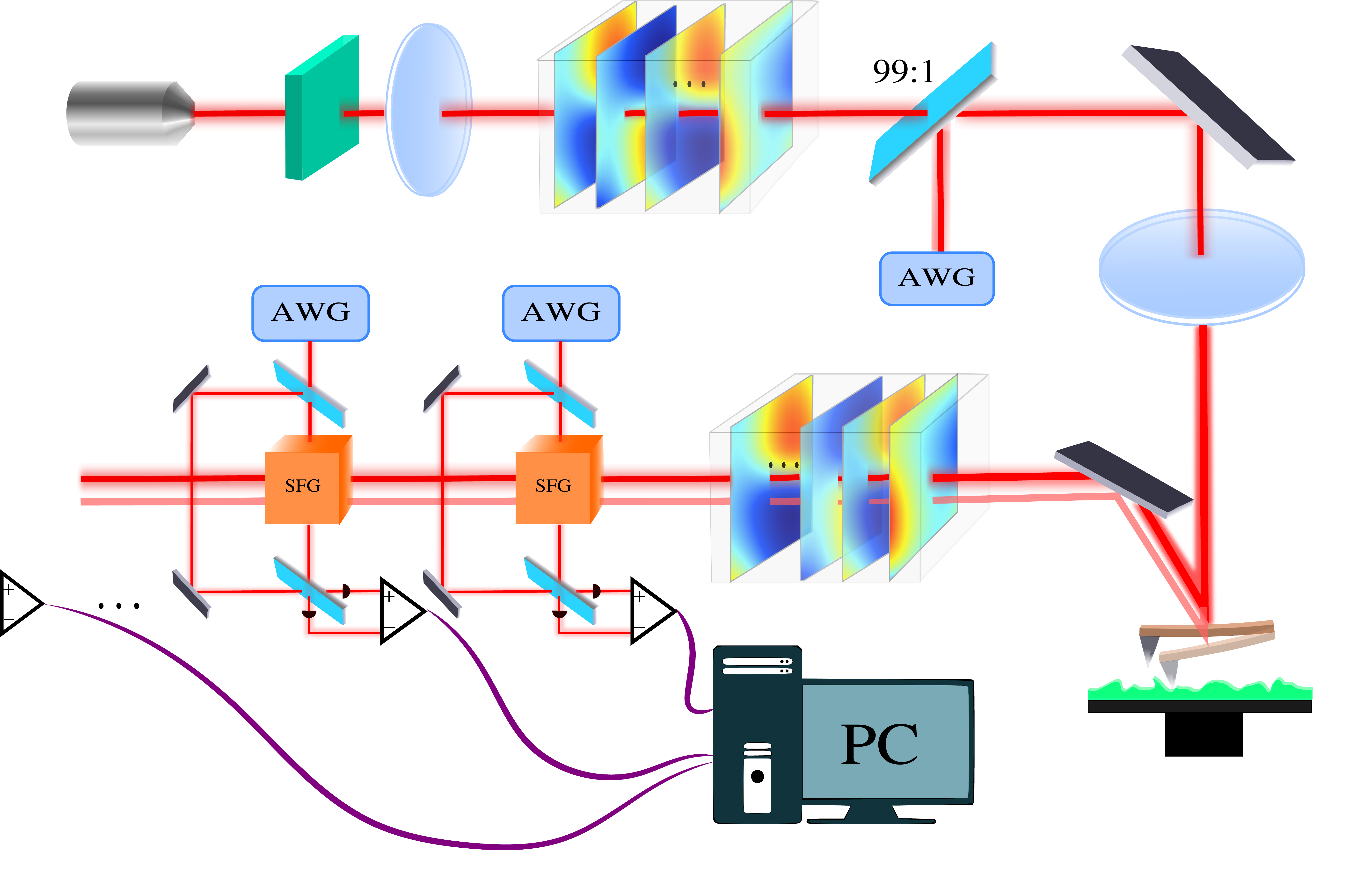}
	\caption{A schematic of the Gaussian multi-mode-entangled transmitter and receiver, depicted in an atomic force microscopy (AFM) setup. An optical parametric amplifier (OPA) built using a non-linear crystal is pumped by a continuous-wave laser to produce a multi-spatial-mode squeezed light. Then, a multi-spatial-mode transformation is implemented on it by a stratified propagation through phase spatial-light modulators (SLMs) separated by small Fresnel propagation segments through an isotropic medium. This transformation combines the effects of beamsplitter array in Fig.~\ref{fig:GaussianCircuit} and the unitary $\hat{T}^\dagger$. This is followed by a multi-mode displacement (due to the injected coherent states in Fig.~\ref{fig:GaussianCircuit}), realized by mixing the multimode squeezed light with an appropriately modulated strong laser local oscillator on a high-transmissivity beamsplitter, which could be generated by an arbitrary waveform generator (AWG). This entangled probe is then reflected from the back of a cantilever, the tip of which is probing, for example, the surface of a biological sample \cite{eaton2010atomic}. To estimate the beam displacement due to the movement of the cantilever, stratified propagation through another set of SLMs is used to first apply the multimode unitary $\hat{T}$. Finally, a sequence of mode-selective upconversion, implemented via sum-frequency generation (SFG) pumped by an LO in the desired mode, is used to selectively extract each mode and is homodyne detected. The classical outcome of the homodyne array is processed by a computer to estimate the beam displacement, which maps to a small longitudinal displacement of the cantilever. This procedure is repeated while the probe beam is raster scanned on the sample, in order to map out its surface structure.}
	\label{fig:fullGaussian}
\end{figure}

The procedure to generalize the above spatially-entangled Gaussian transceiver construction to entangled spatio-temporal modes is similar to what we did in the last section. In this case, the block-diagonalized unitary is given by repeating the MZI-array shown in Fig.~\ref{fig:GaussianCircuit} $M_T$ times. The energy distributions for the coherent states stay the same for each temporal mode index, i.e., $c_k=\lambda_k/\sum_{k=1}^{M_SM_T/2}\lambda_k$. Therefore, the estimator in Eq.~\eqref{eq:estimator} remains the same, with the upper limit of summation being extended from $M_S/2$ to $M_TM_S/2$ and $N_S$ being replaced by $N$. Same as when we considered non-Gaussian optimal states, $A$ is invariant under this extension for the eigenvalues $\brac{\lambda_k}$ are periodic. At last, we just need a $M_TM_S/2$ mode beamsplitter array, such that $
\hat{a}_0' =\sum_k \sqrt{c_k}\hat{\gamma}_k^s
$, to entangle across all spatial-temporal modes. Putting a squeezed vacuum in mode $\hat{a}_0'$ with average photon number $N_s=N/2$, we get the same minimum estimation error as in Eq.~\eqref{eq:entangled-rms}.

Finally, let us discuss how one might assemble a transceiver structure for the entangled Gaussian transmitter developed in this section. A notional schematic is shown in Fig.~\ref{fig:fullGaussian}. The transmitter generates multi-spatial-mode squeezed light using an optical parameter amplifier (OPA) with a known modal squeezing content~\cite{kwon2017single}, which is then transformed into the desired spatially-entangled squeezed state using a universal volumetric mode sorter. Many physical realizations of spatial mode-transformation devices have been explored in the literature. One of those, which uses a stratified free-space propagation through an isotropic medium interspersed with phase masks (which can be realized for example with spatial light modulators or deformable mirrors)~\cite{morizur2010programmable} can in principle realize arbitrary multi-spatial-mode transformations~\cite{borevich1981description}. The multimode quadrature displacement caused due to the injected coherent states shown in Fig.~\ref{fig:GaussianCircuit} is realized by mixing the multi-mode entangled squeezed light with an appropriately phase-and-amplitude modulated high-intensity local oscillator laser on a highly-transmissive (e.g., 99:1) beam splitter ~\cite{paris1996displacement}.  At the receiver, we need another volumetric spatial mode transformation followed by a succession of mode-selective homodyne detections, which can in turn be implemented using a quantum-state-preserving mode-selective frequency up-conversion of one mode at a time using non-linear sum-frequency generation (SFG)~\cite{manurkar2016multidimensional}. The generalization of this architecture to spatio-temporal entanglement follows in a straightforward way. The entire architecture can in principle be realized with available technology.

\section{Discussion and Conclusion}\label{sec:conclusion}

We establish the ultimate quantum limit of the accuracy with which one can detect a small lateral movement of an optical beam. We find an explicit recipe of how to optimally excite the spatio-temporal modes in the time-bandwidth-space product afforded by the probe's duration, optical bandwidth and propagation geometry. That optimal spatio-temporally-entangled probe can be generated by a single-mode continuous-wave squeezed light source along with a universal spatial-mode mixer~\cite{morizur2010programmable} and a universal temporal-mode mixer~\cite{Lukens2017-fv}. This quantum optimal probe exhibits a curious super-Heisenberg scaling in terms of the number of spatial modes excited. Since the production, transformation and detection of Gaussian quantum states (multi-mode squeezed states of light) is far less demanding that non-Gaussian states of light such as N00N states, our scheme is much more feasible to realize in the near term compared to other applications of photonic quantum enhancements, such as universal photonic quantum computing~\cite{Pant2019-ds} or all-optical repeaters for entanglement distribution~\cite{Pant2017-mp}.

In any real system, there will be losses intrinsic to the sources and within the receiver. The latter may include coupling inefficiency of the collected light, the detector's intrinsic inefficiency, and mode-sorting inefficiency. There might also be scattering, absorption and turbulence in free-space propagation path depending upon the application scenario. When loss is taken into account, an analysis along the lines of that done in Ref.~\cite{zhuang2018distributed} will have to be done, and will be an important direction for follow-on work. We expect the high ${\bar n}$ scaling of $\delta d$ to be same for all three sensor types: classical, spatial-only entangled, and spatio-temporal-entangled. However, we expect constant-factor gaps in the high $\bar n$ regime, that is a function of the loss, and the number of spatial and temporal modes (the effective number of ``distributed sensors"). Although photon losses is detrimental to quantum advantage in sensing in general, for entanglement-assisted distributed sensing, when the number of sensors (for our problem, the number of spatio-temporal modes) $M$ is large, an appreciable performance gain can still be obtained for moderate losses as shown in Ref.~\cite{zhuang2018distributed}.

In our analysis of the sensor performance in the diffraction-limited near-field vacuum-propagation path, we took the first $D$ modes to be unit transmissivity, where $D$ is the free-space Fresnel number product, and all higher modes to have zero transmissivity. This sharp drop off of modal transmissivity is an excellent approximation in the near-field regime, especially for hard-aperture pupils (see Fig. 3 in~\cite{holevobook}). Ignoring the lossy higher-order modes is justified since we may simply choose not to excite those modes. Once losses have been incorporated in our analysis, (1) our results will naturally extend into the far-field regime, and (2) in the near field regime, we will get a slight improved performance than what our results suggest, since a few additional lossy modes (beyond the $D$-th mode) will contribute to the beam displacement estimation.

In biological imaging applications, it is important to get a high quality image while ensuring the cellular processes being investigated are in their {\em in vitro} state \cite{cole2014live}, which imposes a constraint on the probe illumination power. Since our scheme can obtain a desired accuracy with less illumination power compared with a classical probe, and since spatial-entanglement enhancement is possible only in the diffraction-limited near-field regime, our results could be extremely relevant for biological imaging applications such as molecular tracking or cellular imaging~\cite{taylor2013biological,taylor2014subdiffraction}. The low probe power also makes this scheme attractive for covert sensing~\cite{bash2017fundamental} where the goal of the sensor is to prevent the detection of the optical probing attempt by an adversary by hiding the probe signal within the thermal noise floor.

Even though the analysis in this paper was for a one-dimensional setting, generalizing to two dimensional (i.e., vector) displacements is straightforward. One interesting direction of future work would be to generalize our results on estimation of a (given, constant) beam displacement to the precision of tracking of a (temporally-varying) beam displacement. Another intriguing future direction is to study quantum enhancements in sensing both transverse and longitudinal movement of an optical beam, with applications to vibrometry, doppler ranging, and 3D imaging. 

\section*{Acknowledgements}

SG thanks Zheshen Zhang for introducing him to Ref.~\cite{zhuang2018distributed}, and Linran Fan for helpful comments on the manuscript. SG acknowledges funding from the Office of Naval Research. HQ is grateful to Zachary Vernon, Casey Myers, Krishna Sabapathy and Daiqin Su for helpful discussions.

\section*{Disclosures}
The authors declare no conflicts of interest. This document does not contain technology or technical data controlled under either the U.S. International Traffic in Arms Regulations or the U.S. Export Administration Regulations.

\bibliography{displacement_ref}

\clearpage

\appendix

\section{Classical and quantum Fisher information}\label{app:FisherInfo}

Consider the problem of estimating a parameter $d$ encoded in a quantum state $\rho_d$ by making a suitable joint measurement on $\nu$ independent copies of $\rho_d$. A quantum measurement on one copy of $\rho_d$, described by positive-operator valued measurement  $\brac{\Lambda_m}$, produces a measurement outcome $m$ with probability distribution $p(m;d) = \tr\brac{\rho_d \Lambda_m}$. Assuming the same measurement is performed on all $\nu$ copies of $\rho_d$, the minimum error of estimating $d$ (from $\nu$ i.i.d. samples of $m$ drawn from the distribution $p(m;d)$) using an unbiased estimator is lower bounded by the inverse of the {\em classical Fisher information} (CFI), $\F_C(p) = \int d{m} \,p({m;d})\frac{\partial^2}{\partial d^2}{\ln p({m;d})}$, also known as the {\em Cram\'{e}r-Rao bound}. In other words,
\begin{align}\label{eq:CCRB}
\delta d^2 \geq \frac{1}{\nu\mathcal{F}_C(\rho_d,\brac{\Lambda_m})},
\end{align}
If we optimize this classical Cramer-Rao bound over all possible measurement choices, the ultimate error of any unbiased estimator of the displacement $d$, is given by the {\em quantum Cram\'{e}r-Rao bound} \cite{braunstein1994statistical,braunstein1996generalized}, which in general gives a tighter bound compared to classical Cram\'{e}r-Rao bound corresponding to any specific measurement $\brac{\Lambda_m}$:
\begin{align}\label{eq:QCRB}
\delta d^2 \geq \frac{1}{\nu\mathcal{F}_Q(\rho_d)} \geq \frac{1}{\nu\mathcal{F}_C(\rho_d,\brac{\Lambda_m})}, \, \forall \brac{\Lambda_m}.
\end{align}
Here, $\F_Q$ is called the {\em quantum Fisher information} (QFI), which is a function just of $\rho_d$, i.e., calculating the QFI does not require us to specify a measurement. Specifically, the QFI is given by the following expectation value,
\begin{align}\label{eq:QFI-formula}
\F_Q(\rho_d) = \Tr\brac{\rho_dL(\rho_d)^2}~,
\end{align}
where the Hermitian operator $L(\rho_d)$ is the so-called 
{\em symmetric logarithm derivative operators} (SLD). When written in the eigenbasis of state $\rho_d=\sum_{i}\lambda_i(d)\ket{\lambda_i(d)}\bra{\lambda_i(d)}$, the SLD explicitly reads:
\begin{align}
L(\rho_d) = \sum_{i,j}\frac{2\bra{\lambda_i(d)}\dot{\rho}_d\ket{\lambda_j(d)}}{\lambda_i(d)+\lambda_j(d)}\ket{\lambda_i(d)}\bra{\lambda_j(d)}~,
\end{align}
where the sum takes over all non-vanishing eigenvalues. Just like CFI, the QFI defined above is also additive, i.e., $\F_Q(\rho_d^{\otimes N})=N\F_Q(\rho_d)$.
It was further shown that the quantum Cram\'{e}r-Rao bound can always be saturated asymptotically by maximum likelihood estimation and a projective measurement in the eigenbasis of the SLD~\cite{braunstein1994statistical,braunstein1996generalized}. 

A particular useful and relevant formalism for us is the QFI of the output state resulting from a unitary evolution of a pure input state, $\ket{\psi_d} = e^{i\hat{H}d}\ket{\psi}_{\text{in}}$. In this case, Eq.~\eqref{eq:QFI-formula} reduces to 
\begin{align}
\F_Q(\ket{\psi_d}) = 4 \sbrac{\bra{\psi_d}\hat{H}^2\ket{\psi_d}-\abs{\bra{\psi_d}\hat{H}\ket{\psi_d}}^2}~.
\end{align}

For the problem being considered in this paper, we are aiming at finding the optimal input (probe) state that results in a modulated state $\rho_d$ with the highest QFI. Therefore, it suffices for us to just consider pure input states, thanks to the convexity of QFI~\cite{fujiwara2005quantum}~\footnote{According to the Schimdt decomposition, any mixed state can be decomposed into a mixture of pure states: {$\rho =\sum_x p(x)\ket{x}\bra{x}$}. The output state  is thus given by {$\rho(\theta)=\sum_x p(x) U(\theta)\ket{x}\bra{x}U^\dagger(\theta)$}. Applying the concavity of QFI, we have {$\max_{\rho}\F_Q(\rho)\leq \max_{\ket{x}}\F_Q(U(\theta)\ket{x})$}}.
For a unitary of the form $\exp[i\tilde{d}\hat{H}]$, ${\tilde d} = d/r_R$ and a pure input state, the QFI is independent of $d$ and is given by following quantity, proportional to the variance of the Hamiltonian $\hat{H}$:
$
\F_Q = \frac{4}{r_R^2}\pbrac{\ave{\hat{H}^2}-\ave{\hat{H}}^2}~.
$

\section{Hermite-Gaussian modes}
In this section we review and derive some basis properties of Hermite-Gaussian modes. We first define {\em Hermite function} of order $n$:
\begin{align}
\psi_n(x) =\ (2^nn!\sqrt{\pi})^{-\frac{1}{2}}e^{-x^2/2}H_n(x)~.
\end{align}
Then the Hermite-Gaussian modes with waist size $w_0$ are given by
\begin{align}
u_n(x) = \pbrac{\frac{2}{w_0^2}}^{\frac{1}{4}}\psi_n\pbrac{\frac{\sqrt{2}x}{w_0}}~.
\end{align}
The matrix $\mb{\Gamma}_{mn}$ is central to our calculation, which is given by
\begin{align}
\mb{\Gamma}_{mn} &= \int_{-\infty}^\infty dx u'_m(x)u_n(x),\\
&=\frac{\sqrt{2}}{w_0}\intinf \psi'_m\pbrac{x}\psi_n(x),\\
&=\frac{\sqrt{2}}{w_0}\intinf \sbrac{\sqrt{\frac{m}{2}}\psi_{m-1}(x)-\sqrt{\frac{m+1}{2}}\psi_{m+1}(x)}\psi_n(x),\\
& = \frac{1}{w_0}\sbrac{\sqrt{m}\delta_{m-1,n}-\sqrt{m+1}\delta_{m+1,n}}\label{eq: Gamma-supp}.
\end{align}
Therefore, only nearest-neighbour coupling exists. 
\section{Single-mode probe cannot beat SQL}
Here we consider an arbitrary choice of othornormal modes $\brac{u_n(x)}$ with annihilation operators $\brac{\hat{a}_n}$. Without loss of generality, consider a single-mode state on the zero mode as follows,
\begin{align}
\ket{\psi}\otimes\ket{0\cdots 0\cdots}~.
\end{align}
The calculation of the variance $\Delta^2\hat{H}$ is straightforward. For the expectation value we have
\begin{align}
\ave{\hat{H}} &= -i\sum_{mn}\ave{\mb{\Gamma}_{mn}\hat{a}^\dagger_m\hat{a}_n}\\
& = -i\mb{\Gamma}_{00} \bra{\psi}\hat{a}^\dagger_0\hat{a}_0\ket{\psi} = 0~.
\end{align}
We thus have $\Delta^2\hat{H} = \ave{\hat{H}^2}$, which is equal to
\begin{align}
\ave{\hat{H}^2}& = -\sum_{mn}\sum_{kl}\mb{\Gamma}_{mn}\mb{\Gamma}_{kl}\ave{\hat{a}^\dagger_m\hat{a}_n\hat{a}^\dagger_k\hat{a}_l}\\
& = -\sum_{n,k}\mb{\Gamma}_{0n}\mb{\Gamma}_{k0}\ave{\hat{a}^\dagger_0\hat{a}_n\hat{a}^\dagger_k\hat{a}_0}\\
& = -\sum_{n=0}^\infty \mb{\Gamma}_{0n}\mb{\Gamma}_{n0}\bra{\psi}\hat{a}_0^\dagger \hat{a}_0\ket{\psi}\\
& = \mb{\Gamma}_{0n}\mb{\Gamma}_{0n}^* N_S~.
\end{align}
Using the completeness relation $\sum_nu_n(x)u_n^*(x')=\delta(x-x')$, we have
\begin{align}
\delta d = \frac{C}{2\sqrt{N}}
\end{align}
where 
\begin{align}
1/C^2 = \int dx \abs{\frac{\partial u_0(x)}{\partial x}}^2~.
\end{align}
Therefore, we have prove that single-mode state cannot beat SQL for any choice of othornormal modes, and higher-order spatial modes gives better sensitivity. 
\section{Asymptotic behaviour of the Schatten norm of the coupling matrix }
As shown above, coupling matrix for HG modes is given by the following skew-symmetric matrix 
\begin{align}
\mb{\Gamma}(M)=
\begin{bmatrix}
0 & 1 & 0 & \dots & 0\\
-1 & 0 & \sqrt{2} & \dots & 0\\
0 & -\sqrt{2} &  0 &\ddots &\vdots \\
\vdots & \vdots & \ddots & \ddots &\sqrt{M-1} \\
0 & 0 &\dots & -\sqrt{M-1}&0
\end{bmatrix}~,
\end{align}
The eigenvalues of any skew-symmetric matrix are imaginary pairs $\brac{\pm i\lambda_k}$ where $\lambda_k>0$. Therefore the sum of power of $\lambda_k$ is related to the schatten norm
\begin{align}
\sum_i \lambda_i^p =\frac{1}{2}\vert\mb{\Gamma}\vert_p^p~. 
\end{align}
From the main text, to calculate the QFI of optimum entangled probe we need to know the sum for $p=2$ and $p=1$ in the limit $M\gg 1$. The former is straightforward to calculate :
\begin{align}
\sum_k\lambda_k^2 &= -\frac{1}{2}\tr\brac{\mb{\Gamma}^2}\\
& = -\frac{1}{2}\sum_{nm}\mb{\Gamma}_{mn}\mb{\Gamma}_{nm}\\
& = \frac{1}{2}\sum_{n=1}^{M-1}n = \frac{M(M-1)}{4}~,
\end{align}
where we use $\mb{\Gamma}_{mn}$ for HG modes in Eq.~\eqref{eq: Gamma}.

The calculation when $p=1$ is more complicated. We first observe that the symmetric version of $\mb{\Gamma}$
\begin{align}
\tilde{\mb{\Gamma}}(M)=
\begin{bmatrix}
0 & 1 & 0 & \dots & 0\\
1 & 0 & \sqrt{2} & \dots & 0\\
0 & \sqrt{2} &  0 &\ddots &\vdots \\
\vdots & \vdots & \ddots & \ddots &\sqrt{M-1} \\
0 & 0 &\dots & \sqrt{M-1}&0
\end{bmatrix}~,
\end{align}
has eigenvalues $\brac{\pm\lambda_i}$. Thus, for our purpose it is sufficient to consider the symmetric matrix above. Next, we observe that the characteristic polynomial of $\tilde{\mb{\Gamma}}$ is proportional to the Hermite polynomial, $2^{-M/2}H_M(\lambda/\sqrt{2})$. Arranging $\brac{\lambda_k}$ such that they are decreasing, the asymptotic form of $\lambda_k$ is given by~\cite{dominici2007asymptotic}
\begin{align}
\begin{cases}
\lambda_k \simeq \sqrt{M} +O(M^{1/6}),~~ \text{for}~~ k=O(1)~,\\
\lambda_k \simeq \frac{\pi}{4}(M-2k)\sbrac{M^{-1/2}+O(M^{-3/2})}, ~~ \text{for}~~ k=O(n/2)~.
\end{cases}
\end{align}
From which we can see that $\sum_k\lambda_k \simeq M^{3/2}$.

\end{document}